# From individual to strongly coupled metallic nanocavities

Adi Salomon and Yehiam Prior

*Department of chemical physics, Weizmann Institute of Science, Rehovot, Israel*

[1,2]Radoslaw Kolkowski and [1]Joseph Zyss

[1]*Laboratoire de Photonique Quantique et Moleculaire, Institut d'Alembert, Ecole Normale Supérieure de Cachan, France*

[2]*Institute of Physical and Theoretical Chemistry, Faculty of Chemistry, Wroclaw University of Technology, Poland*

Localized plasmonic modes of metallic nanoparticles may hybridize like those of atoms forming a molecule. However, the rapid decay of the plasmonic fields outside the metal severely limits the range of these interactions to tens of nanometers. Herein, we demonstrate very strong coupling of nanocavities in metal films, sparked by propagating surface plasmons and evident even at much larger distances of hundreds of nanometers for the properly selected metal/wavelength combination. Such strong coupling drastically changes the symmetry of the charge distribution around the nanocavities making it amenable to probing by the nonlinear optical response of the medium. We show that when strongly coupled, equilateral triangular nanocavities lose their individual three-fold symmetry to adopt the lower symmetry of the coupled system and then respond like a single dipolar entity. A quantitative model is suggested for the transition from individual to strongly coupled nanocavities.

When two or more metallic nanoparticles are in close proximity, their plasmonic modes may interact through the near field, leading to additional resonances of the coupled system or to shifts of their plasmonic modes resonant frequencies.[1-10] This process is analogous to atom-hybridization, as had been proposed by Gersten and Nitzan[11] and modeled by Nordlander and Halas[12]. The coupling between the plasmonic modes can be in-phase (symmetric) or out-of-phase (anti-symmetric), reflecting, correspondingly, the 'bonding' and 'anti-bonding' nature of such configurations. [13-15] Since the incoming light modifies the charge distribution around the metallic nanoparticle, its polarization plays a major role in shaping the energy-level landscape upon hybridization as was studied by optical linear measurements[13, 16, 17].

Nanocavities (nanoholes in a thin metal film) are structures which are complementary to nanoparticles and their linear optical response may be described through Babinet's principle[18]. However, when considering interaction between neighboring structures, a fundamental difference exists between nanocavities and nanoparticles. While the near field interaction between nanoparticles vanishes rapidly (typically within tens of nanometers), propagating surface plasmons polaritons (SPP) which are supported by the metal film offer an additional mechanism for coupling between the nanocavities that fundamentally changes the picture.

The coupling between metallic nanostructures can be elucidated by nonlinear, polarization sensitive optical measurements. In particular, the dependence of Second Harmonic Generation (SHG) on the polarization of the incident beam provides direct information about the properties of the second order susceptibility tensor $\chi_{ijk}^{(2)}$ which, in turn, reflects the underlying electronic properties and charge distribution at the surface of the metallic nanostructure. [19-21] Furthermore, the nature of the susceptibility tensor is determined by the system symmetry, and thus, if the coupling between individual nanostructures leads to symmetry changes, dramatic changes both in the intensity and the polarization properties of the SHG responses are expected.

To study their coupling, and for reasons which are deeply rooted, we chose to work on equilateral triangular nanostructures (nanocavities or nanoparticles)Three-fold symmetry is the highest level of symmetry, compatible with centro-symmetry breaking, and hence with SHG, Thus, it stands-out as the best starting point to observe symmetry lowering effects due to the . Indeed, this unique choice provide striking differences between the SHG response of an individual structure which naturally obeys a threefold a-polar symmetry to that of a coupled nanostructures with a lower polar symmetry. Our samples thus consist of equilateral triangular nanocavities in a thin silver film and their complementary structures, triangular silver nanoparticles, all of typical side length of 200nm (see figs 1a,b). The triangular cavities/particles are arranged in a linear configuration separated by a distance of about 400 nm (figs 2a,e). When strongly coupled, the overall symmetry of the SHG response is reduced to a typical twofold dipolar shape, that is a single mirror perpendicular to the substrate and containing the coupling axis, as opposed to three mirrors in a three-fold symmetrical configuration for an individual structure.

First we study the nonlinear responses of individual nanostructures[22]. The total SHG emission intensity patterns for both triangular nanoparticle and nanocavity structures appear to be reasonably independent of the input beam polarization and they display a similar behavior as is shown in Fig. 1.

Next we study the nonlinear response of a linear array of three triangular nanocavities and of the complementary set of nanoparticles when excited by linearly polarized light with a polarization axis along, or perpendicular to the common axis. Fig 2 depicts the spatial distribution of the SHG responses for the two triplets and for the two orthogonal input polarizations. Figs 2.b,c shows, for the nanoparticles, the spatial distribution of the SHG intensity for excitation along (perpendicular to) the triplet axis. In both cases the intensity seems to be almost uniformly spread, and there is very little difference between the excitation by the two input polarizations of the incoming beam. For the nanocavities (Fig 2f,g), however, the outcome is very different, and a dramatic change in the SHG responses is observed depending on whether the input polarization is pointing along or perpendicular to the triplet axis. When the input polarization is along their common axis (the "Y-axis"), not only is the SHG intensity much higher, but it is also spatially concentrated near the center of the structure. To further characterize the SHG emission from these structures, we measured their polarization dependencies. The **_total_** SHG intensity is polar-plotted against the input beam linear polarization angle for both systems. The nanoparticles (fig 2d) display a typical almost uniform signature, with similar intensities for all input polarization directions (fig 2d). For, the nanocavities however (fig 2h), the nature of the emission has completely changed compared to that of an individual cavity and a clear dipolar signature is evident, with a much stronger intensity (by a factor of 4) when excited by light polarized along the common Y axis. The detailed experimental results and a comparison to theory are presented below. To confirm that we measure only the SHG, and attest that any contribution of luminescence or fluorescence from impurities are totally negligible[23], we measured the spectra of the emitted light both for the nanoparticles and the nanocavities as shown in Fig 3. In both cases the SHG signal at twice the input beam frequency is the only component present in the spectra. Fig 3a shows the SHG spectrum emitted from triangular nanoparticles excited by two orthogonal polarizations. One notices that in this case both polarizations give similar intensities. In the case of the nanocavities, however, the situation is, again, very different– the SHG intensity is higher by factor ~5 when excited along the common Y-axis (Fig 3b), in agreement with the results for the spatial distribution shown above.

The observed SHG responses of the two structures indicate that different physical mechanisms are in operation for each case. Whereas the nanoparticles are responding independently and their excitation is insensitive to the polarization of the input beam, the nanocavities respond as a single entity with a much stronger, spatially concentrated and directed SHG emission. When coupled, their original threefold symmetry is lowered to twofold symmetry, resulting in a strongly dipolar emission pattern. In what follows we perform a direct analysis of the symmetry properties of such structures, and provide a very simple tool to analyze the degree of coupling between the nanocavities. A more detailed analysis is provided in the supplementary material, and a full account of the theory will be published elsewhere.

To a good approximation our samples can be viewed as two-dimensional, and for incoming transverse polarizations, the contribution from the Z component (perpendicular to the plane of the thin film) of the electric fields may be neglected. The dominant contributions to the susceptibility tensor $\chi_{ijk}^{(2)}$ span only the Y and X directions (in the film's plane), leaving $\chi_{YYY}^{(2)}$, $\chi_{YXX}^{(2)}$, $\chi_{XXX}^{(2)}$ and $\chi_{XYY}^{(2)}$ as the only four independent coefficients. The presence of an additional mirror plane normal to the X axis leads to the cancellation of tensor elements which are anti-symmetric with respect to this mirror symmetry, thus, terms where the X index appears an odd number of times vanish, namely $\chi_{XXX}^{(2)} = \chi_{XYY}^{(2)} = 0$. For a higher symmetry corresponding to three-fold rotational invariance around Z, such as for our equilateral triangular structures, the two remaining coefficients are further linked by the relation:

$$\chi_{YYY}^{(2)} = -\chi_{YXX}^{(2)} \tag{1}$$

which leaves only one independent coefficient [24]. This relation expresses the cancellation of dipolar-like entities attached to a trigonal structure, the $\chi_{YYY}^{(2)} + \chi_{YXX}^{(2)}$ combination of tensor coefficients behaving precisely as the Y component of such a dipolar quantity.

If, on the other hand, the strong coupling between the triangular nanocavities reduces the threefold symmetry to a twofold reflection symmetry with respect to the Y axis, relation (1) no longer holds, and the $\chi_{YYY}^{(2)}$ and $\chi_{YXX}^{(2)}$ terms become independent again.

According to these very general symmetry considerations, the level of coupling between the nanocavities can be monitored via the dependence of the emitted SHG signal on the input beam polarization. As can be seen in Fig 2, for strong coupling, when the input polarization is along the Y-axis, the three cavities respond collectively (Fig. 2g) and the polarization signature of the emitted light (fig 2h) changes from the characteristic four-fold octupolar shape, characteristic of a triangle, [22, 25] to an elongated dipolar signature characteristic of a reduced, two-fold reflection symmetry. We define a nonlinear anisotropy coefficient ρ, which reflects the departure from three-fold symmetry due to the coupling between the metallic nanocavities:

$$\rho = \chi_{YXX}^{(2)} / \chi_{YYY}^{(2)} \tag{2}$$

This parameter is physically meaningful and practical as it allows to quantitatively monitor the transition from three independent objects to a single internally coupled entity of lower symmetry. As the level of coupling between individual nanostructures increases, the strong SPP induced coupling between the nanocavities gives rise to the observation of the entire system behaving as a single unit (dipole). In such a case, $\chi_{YYY}^{(2)}$ tends to dominate over $\chi_{YXX}^{(2)}$, ρ ultimately approaching 0. On the other hand, in the absence of coupling between the individual nano-structures, the system maintains its original three fold symmetry, relation (1) holds and ρ=−1. These two extreme cases of individual contributions ($\rho=-1$) and the fully coupled one ($\rho=0$), calculated under realistic experimental conditions, are shown in fig 4. As can be clearly seen, for $\rho=-1$, the SHG emission features octupolar patterns for both the X and Y components and the total emission (black line) is barely sensitive to the incoming beam polarization. [22] The emission pattern in this case, is very similar to that of a single triangular cavity/particle or to that of triplet of triangular nano-particles which are barely coupled (fig 2d). For $\rho=-0.1$, a dipolar emission is observed for both X and Y components (fig 4b), reflecting the two-fold symmetry of a strongly coupled system, where the threefold symmetry nanostructures act in unison as a single dipole, locked along the same axis and emitting in phase.

To investigate the degree of coupling between the triangular nanocavties, we performed detailed polarization analysis of the individual X- and Y components of the emitted SHG radiation as shown in Fig 5. A dipolar-like pattern is observed for the polarization along the Y-axis (blue curve); while an octupolar pattern of weaker intensity is observed along the X-axis (red curve). The dipolar pattern results from the reduced symmetry due to the strong coupling between the LSP modes. However, the coupling is not fully accomplished and the X component emission pattern features an octupolar pattern [22, 25] as is expected from our nonlinear anisotropy model. Indeed, the data was best fitted to the model described above when $\rho=-0.5$. This value indicates a relatively strong, however still intermediate level of coupling between the nanocavities.

We note that under identical conditions, the triangular nanoparticles are not coupled at all, portraying octupolar signature typical of individual triangles.

To summarize, when it comes to coupling between neighboring metallic nanostructures, a fundamental difference exists between nanocavities and nanoparticles, and the two structures cannot be considered as exactly complementary. Unlike nanoparticles, nanocavities in metal films are coupled through excitation of SPP [26]. These SPP modes connect localized surface plasmons (LSP) at neighboring nanocavities, inducing strong mutual interactions. The degree of coupling should depend on the SPP excitation wavelength and its propagation length and therefore on distance between the cavities. The intrinsic parameters of the metallic film itself should play a role as well. For example, poor coupling was observed (not shown here) also when the metallic surface between the cavities was rough leading to scattering of the SPP modes, phase randomization of the dipoles, and therefore reduced coupling efficiency between the LSP modes. A high fabrication quality of the silver film (roughness and grain size) and of the structures themselves proved to be essential for our observations. Furthermore, when gold was used instead of silver, no coupling was observed since gold does not support plasmonic modes at the SHG frequency of the current experiments. Improvement of the coupling between the cavities may be achieved by optimizing the distance between them and the wavelength of the input beam. The understanding of the coupling between neighboring nanoparticles and nanocavities will prove essential for the introduction of functionalities into complex nano structures.

*Methods*

The metallic nanostructures (cavities and particles) were fabricated by a focused ion beam machine (*FEI, Helios Nano Lab 600i*) in a 200nm thick silver film (see figs). The film was evaporated onto a clean fused silica glass under high vacuum conditions; its roughness and grain size were measured (Atomic force microscopy and Scanning electron microscopy) to be smaller than 1nm and 50nm respectively. To have similar indices of refraction from both sided of the nanocavities or around the nanoparticles, the silver surfaces were covered by a 150 nm thick polyvinyl alcohol (PVA) layer with an average refractive index in the visible to near infrared similar to the glass (~ 1.5 ).

Experimental system for the measurement of SHG: The samples were illuminated by a tunable Ti:Sapphire laser (*Spectra-Physics Mai-Tai HP*, 100 fsec, 80MHz, 2-6 mWatt at the entrance lens, with a fundamental incoming beam tunable between 750nm-980nm. The laser was focused through the glass substrate using a 0.7 NA objective (×60), resulting in a spot size of about 0.9 µm. The SH signal was collected in reflection mode through the same objective, and directed to two avalanche photodiodes (APD, PerkinElmer) that measure the SH intensity along the X and Y perpendicular polarization directions. A dichroic mirror was used to block the reflected fundamental beam, and appropriate band-pass filters (*Semrock*) were used to further isolate the SH radiation. For the measurement of the polarization dependence of the emitted SHG, the linear input beam polarization was rotated by a half wave plate.

Acknowledgements:

We acknowledge the support of the Weizmann-CNRS NaBi LEA laboratory. This work was supported by a grant from the Israel Science Foundation. RK acknowledges the government for the Polish-French "co-tutelle" PhD program.

26. Parsons, J.; Hendry, E.; Burrows, C. P.; Auguie, B.; Sambles, J. R.; Barnes, W. L. *Physical Review B* **2009,** 79, (7).

**Figures:**

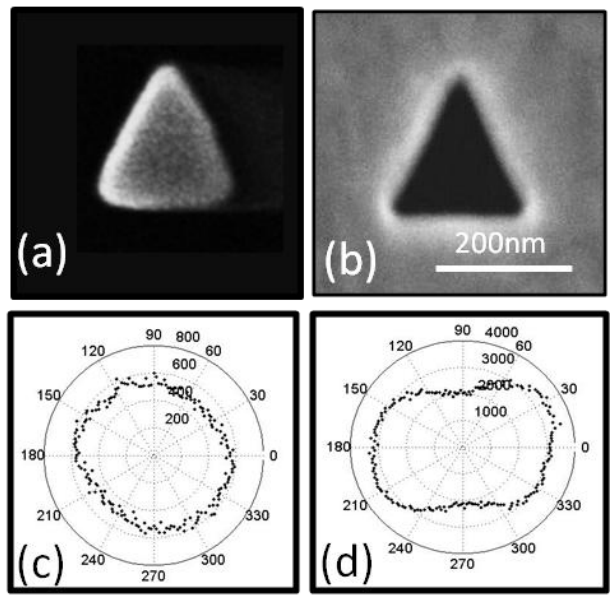

**Figure 1:   SEM images of a triangular nanoparticle and its complementary nanostructure a triangular nanocavity, and their corresponding SHG emission patterns.**   (a) a triangular nanoparticle with side length of ~200nm.  (b) a triangular nanocavity with the same geometrical dimensions. (c-d) are the corresponding SHG total emission patterns as functions of the polarization of the incoming beam.

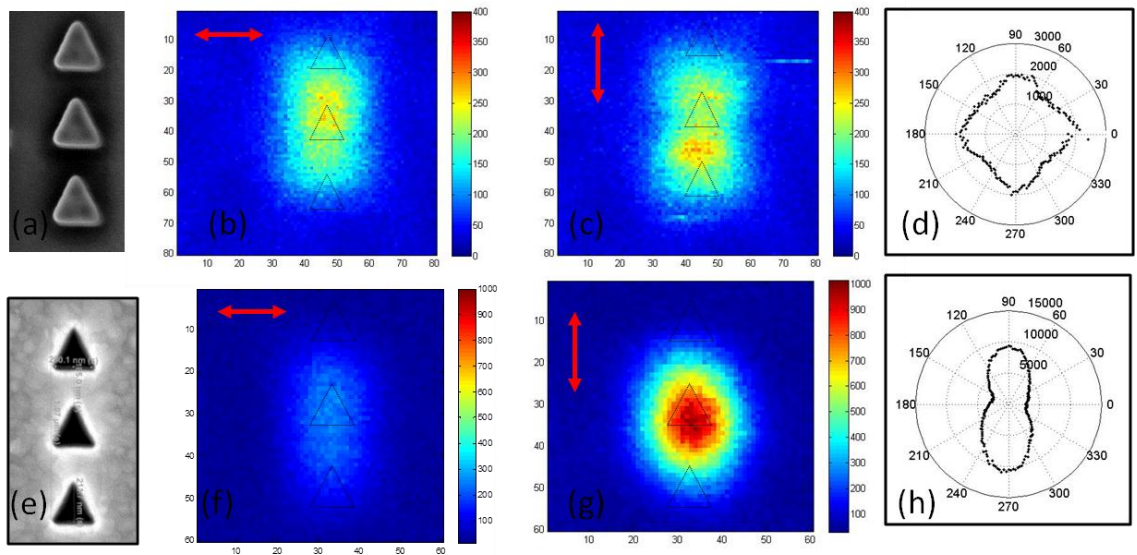

**Figure 2: SHG responses (intensity and emission) as function of the polarization of the incoming beam for triplet of triangular nanoparticles and their complementary nanocavities**. SEM images of (a) triplet triangular nanoparticles with side length of 200nm and separation of ~400nm, and their complementary nanocavities structures (e). SHG scanning of the nano particles with incident polarization along the X-axis(b), and along the Y-axis(c) and their corresponding total emission(d). (f) SHG scanning of the nanocavities with incident polarization along X axis, and along Y axis(g) and their corresponding total emission(h), showing a dipolar-like emission pattern. The fundamental beam wavelength is 940nm.

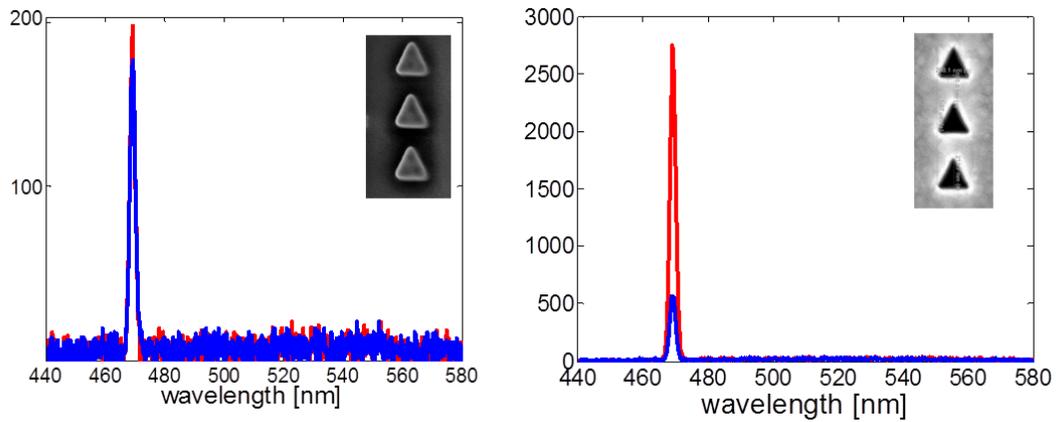

**Figure 3: SHG emission spectra of the metallic nanostructures systems excited with 940nm fundamental wavelength by both X and Y polarizations** (a) Emission spectrum from triplet of triangular nanoparticles; blue curve, incoming beam polarized along the X axis, red – incoming beam polarized along the Y axis. (b) Emission spectrum from triplet of triangular nanocavities; blue curve, incoming beam polarized along the X axis, red – incoming beam polarized along the Y axis.

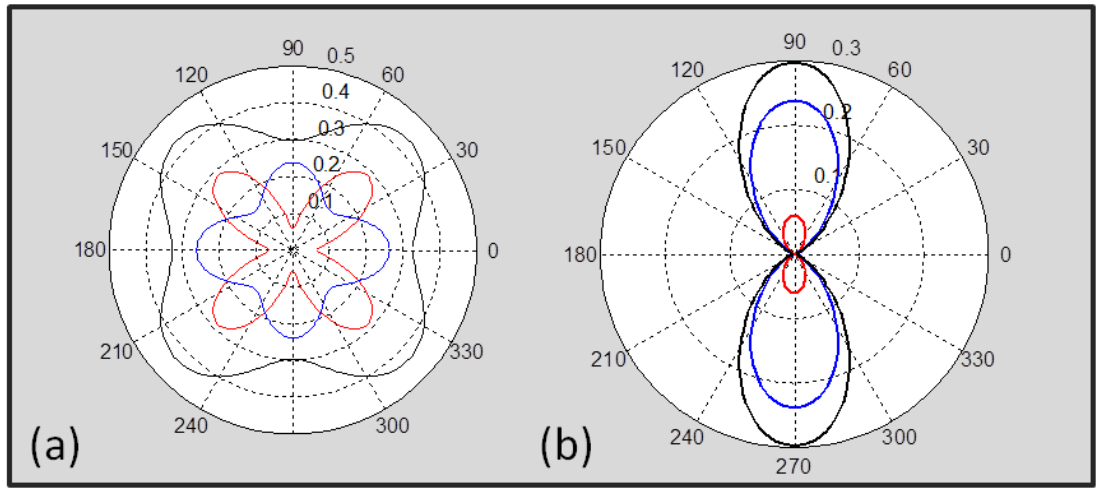

**Figure 4: simulations of the SHG emission pattern as a function of the fundamental beam polarization for objects with three fold symmetry aligned along the Y axis.** The objected are connected through the coupling parameter, ρ. When strong coupling occurs ρ =0 and all the all dipoles are emitting in phase. When ρ =-1, the objects do not interact and the emission is similar to that of an individual object. Blue curve: the Y-axis component collected SHG, red curve: the X-axis component (a) in the absence of coupling between cavities and with randomized phases, or for individual cavity/particle (ρ =-1). (b) when the whole system is strongly coupled and all the emitted dipoles are in phase (ρ =-0.1). The figure includes a small correction for the dichroic mirror which is not fully achromatic.

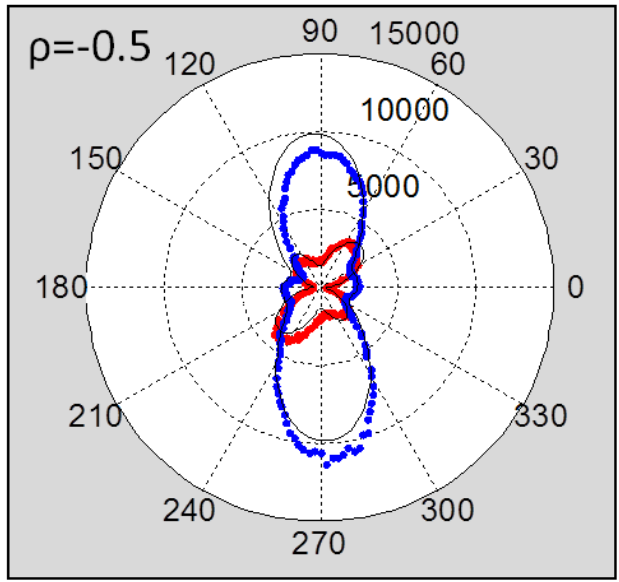

**Figure 5: Fitting the experimental polar plots of the coupled system to the suggested model.** Blue and red dots are the experimental observed polar plots of Y and X components respectively for triplet of triangular nanocavities. The black curve is the fitting of our model with ρ = - 0.5, indicating a strong, however not full level of coupling. The total emission (X+Y) is the one presented in fig2h.